\documentclass[10pt, twocolumn, amssymb, aps, prx, reprint, floatfix, superscriptaddress]{revtex4-2}
\usepackage[a4paper, total={7in, 10in}]{geometry}
\usepackage{graphicx} 
\graphicspath{{Figures/}}
\usepackage{amsmath}
\usepackage{physics}
\usepackage{dsfont}
\usepackage{dcolumn}
\usepackage{xcolor}
\usepackage{bbold}
\usepackage{appendix}
\usepackage{units}
\usepackage{lipsum}
\usepackage{todonotes}

\usepackage{pgfplots}
\pgfplotsset{compat=1.18} 

\usepackage{hyperref}
\hypersetup{colorlinks=true,linkcolor=blue}

\newif\ifptitle
\newif\ifpnumber
\newcounter{para}

\ptitletrue  

\begin{document}

\title{Geometric signatures of the onset of many-body ergodicity}

\author{Chris Ventura-Meinersen}
\thanks{Corresponding author: c.venturameinersen@tudelft.nl}
\affiliation{QuTech and Kavli Institute of Nanoscience, Delft University of Technology, PO Box 5046, 2600 GA Delft, The Netherlands
}

\author{Edmondo Valvo}
\affiliation{QuTech and Kavli Institute of Nanoscience, Delft University of Technology, PO Box 5046, 2600 GA Delft, The Netherlands
}

\author{Stefano Bosco} 
\affiliation{QuTech and Kavli Institute of Nanoscience, Delft University of Technology, PO Box 5046, 2600 GA Delft, The Netherlands
}

\author{Francisco Machado}
\affiliation{QuTech and Kavli Institute of Nanoscience, Delft University of Technology, PO Box 5046, 2600 GA Delft, The Netherlands
}

\author{Maximilian Rimbach-Russ}
\affiliation{QuTech and Kavli Institute of Nanoscience, Delft University of Technology, PO Box 5046, 2600 GA Delft, The Netherlands
}

\date{\today}

\begin{abstract}
    Identifying universal, robust, and interpretable signatures of the onset of many-body quantum ergodicity remains a major challenge. The adiabatic gauge potential has been noted to act as a sensitive probe of quantum chaos.
    Here, we generalize the features of the adiabatic gauge potential to multi-parameter perturbations, yielding an emergent quantum geometry. This new geometry --- whose metric we dub the Hilbert-Killing metric --- provides a tool for probing the onset of ergodicity.
    The Hilbert-Killing metric exhibits a quantifiable fastest growth in the boundary regime, which we corroborate by exact diagonalization of the mixed-field Ising chain and the disordered PXP chain.
\end{abstract}

\maketitle

The emergence of thermalization in quantum many-body systems is, at first glance, at odds with unitary time evolution~\cite{deutschQuantumStatisticalMechanics1991}. 
While unitary quantum dynamics preserves the purity of the global wavefunction, generic interacting systems display rapid relaxation of local observables towards thermal expectation values consistent with highly mixed states~\cite{sierantManybodyLocalizationAge2025}. 
This apparent paradox has motivated extensive efforts to understand when and how complex many-body dynamics give rise to thermal behaviour~\cite{ dalessioQuantumChaosEigenstate2016, abaninColloquiumManybodyLocalization2019}. 
While a variety of experimental studies have explored this question across different quantum simulation platforms (such as neutral atoms~\cite{karchDynamicalPreparationU12026}, superconducting circuits~\cite{abaninObservationConstructiveInterference2025}, Rydberg atom arrays~\cite{geimEngineeringQuantumCriticality2026}, or Fermi-Hubbard simulators~\cite{jirovecManybodyInterferometrySemiconductor2026, farinaSiteresolvedMagnonTriplon2025}), identifying robust and measurable signatures of the onset of quantum ergodicity remains a major challenge.


Previous approaches to diagnose such thermalizing regimes can be divided into three broad classes: 
i) dynamical features, such as the time evolution of local observables, the evolution of entanglement entropy, and operator growth (as probed using out-of-time-ordered correlators~\cite{swingleUnscramblingPhysicsOutoftimeorder2018, abaninObservationConstructiveInterference2025} or scaling of the Lanczos coefficients~\cite{parkerUniversalOperatorGrowth2019});
ii) properties of the eigenstates, such as the size scaling of entanglement entropy~\cite{kimBallisticSpreadingEntanglement2013, torres-herreraDynamicalManifestationsQuantum2017, eisertColloquiumAreaLaws2010}, and the inverse participation ratio (IPR)~\cite{berkeTransmonPlatformQuantum2022};
iii) eigenvalue statistics, such as the gap ratio distribution~\cite{atasDistributionRatioConsecutive2013, tekurHigherorderSpacingRatios2018, sierantModelLevelStatistics2020, jirovecManybodyInterferometrySemiconductor2026}, and the spectral form factor (SFF)~\cite{altlandStatisticsRandomMatrix2025a, liuSpectralFormFactors2018, prangeSpectralFormFactor1997, prakashUniversalSpectralForm2021}. 
While these metrics offer a precise picture deep within thermalizing and non-thermalizing regimes, they are often less well-understood near the boundary, limiting one's ability to identify and characterize it.


\begin{figure}
    \centering
    \noindent\hspace*{-0.8cm}
    \includegraphics[width=1.1\linewidth]{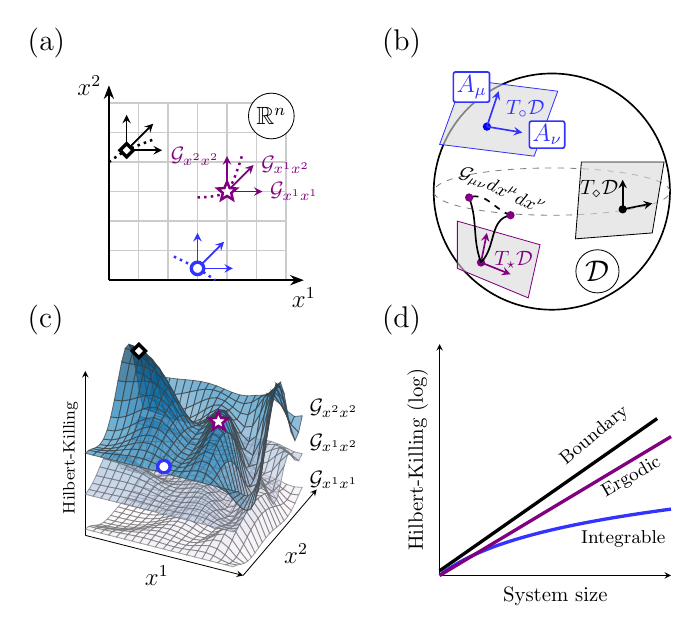}
    \caption{\textbf{Many-body Hilbert-Killing metric.} \textbf{(a)} A system Hamiltonian is defined by a set of parameters, here $(x^1,x^2)$, which corresponds to a possible subset of $n$ parameters. \textbf{(b)} Each point $p\in \mathbb{R}^n$ in the parameter space defines a space of parametrized quantum states $\mathcal{D}$. At a specific point, we can define a (co-)tangent space $T_p \mathcal{D}$ that is spanned by the set of generators of the parallel transport operator $\{A_\mu\}_{\mu=1}^n$. The difference between two infinitesimal parallel transports induces a metric structure, which we call the Hilbert-Killing metric $\mathcal{G}_{\mu\nu}$. \textbf{(c)} In this example, we have two parameters, which induce three components of the Hilbert-Killing metric, which change at each point in the parameter space and which probe the system's response differently depending on the directions $(x^1, x^2)$. In the regime corresponding to the boundary region between integrable and ergodic phases, one finds that the Hilbert-Killing metric has the fastest system size scaling \textbf{(d)}.}
    \label{fig: intro fig}
\end{figure}
Recently, it has been realized that the susceptibility of eigenstates to small perturbations of the Hamiltonian parameters offers a sensitive probe to quantum chaos~\cite{pandeyAdiabaticEigenstateDeformations2020, orlovAdiabaticEigenstateDeformations2023, pozsgayAdiabaticGaugePotential2024, leblondUniversalityOnsetQuantum2021, kimDefiningClassicalQuantum2026}. 
This development rests upon the adiabatic gauge potential (AGP), which encodes the transformation of eigenstates as one perturbs the Hamiltonian. 
In particular, the size scaling of the AGP norm can encode the thermalization regime (ergodic or integrable), as well as sharply identify the boundary between the two~\cite{pandeyAdiabaticEigenstateDeformations2020, pozsgayAdiabaticGaugePotential2024, kimDefiningClassicalQuantum2026, mihailescuSpectrallyLocalGeometric2026}.

One key challenge of the AGP is that its behavior is highly dependent on the nature of the perturbation considered.
Without prior knowledge of the structure of the Hamiltonian and/or its symmetries, one might then mischaracterize the thermalizing regime. 


In this work, we demonstrate how the AGP can be generalized to multi-parameter Hamiltonians, yielding an emerging quantum geometry.
Its metric---which we dub the Hilbert-Killing (HK) metric---encodes how eigenstates deform under changes of the Hamiltonian and serves as a sensitive probe of the onset of ergodicity, Fig.~\ref{fig: intro fig}.
Our results are threefold.
First, we construct the HK metric as a generalization of the AGP~\cite{pandeyAdiabaticEigenstateDeformations2020}, which demonstrates how the AGP, the Killing form, and entanglement fidelity are connected concepts in the study of thermalization, providing a sensitive probe of the onset of ergodicity.
Second, we demonstrate how the interplay between the symmetries of the Hamiltonian and the perturbations leads to disparate behaviors for the HK metric, demonstrating the importance of generalizing the features of the AGP.
Third, we show that for all studied HK metric components, the fastest growth with system size occurs in the ergodic-integrable boundary region in both models. In particular, we identify transient super-exponential scaling as the HK metric approaches the provable upper bound, previously discussed for the single component AGP~\cite{pandeyAdiabaticEigenstateDeformations2020}.

\textit{Entanglement fidelity as Killing metric --- }Consider a Hamiltonian $\hat{H}[x]$ parametrized by $n$ parameters, $x^\mu \in \mathbb{R}^n$ [see Fig.~\ref{fig: intro fig}(a)]. 
For each point $x$, the Hamiltonian $\hat{H}[x]$ determines the eigenstates $\ket{\psi_k}\in\mathcal{D}$.
As one smoothly varies the parameters $x^\mu$, the eigenstates will also change $\ket{\psi_k} = \ket{\psi_k(x)}$.
Crucially, these transformations can be recast as translations in the space of parameterized quantum states of $\mathcal{D}$, where the transformation is given by the AGP along that particular direction~[see Fig.~\ref{fig: intro fig}(b)].
From this point of view, the AGP~\cite{kolodrubetzGeometryNonadiabaticResponse2017}, defined as a Hermitian one-form $A{=}A_\mu dx^\mu$, acts as the generator of the parallel transport operator $U(x_\text{i},x_\text{f}){=}\mathcal{P}\exp\left(-i\int_{x_\text{i}}^{x_\text{f}} A\right)$ with $\mathcal{P}$ referring to the path-ordering operator. 
While the AGP is focused on the transformations along a particular direction in $\mathbb{R}^n$, we can instead consider all possible transformations along $\mathcal{O}(n^2)$ pairs of different directions. 
To this end, let us consider the change of the $k$-th eigenstate under a generic infinitesimal change in parameters $\ket{\psi_k(x{+}dx)}=U(x,x{+}dx)\ket{\psi_k(x)}$.
The infidelity between the two states is strictly non-negative and symmetric, and thus is a well-defined metric on this manifold~\cite{provostRiemannianStructureManifolds1980}:
\begin{align}
    &1-|\braket{\psi_k(x)}{\psi_k(x{+}dx)}|^2 \approx g^{(k)}_{\mu \nu} dx^\mu dx^\nu 
\end{align}
Here the different components of the quantum metric are defined by the adiabatic transformations along different directions, $A_\mu\ket{\psi_k}=i\frac{\partial}{\partial x^\mu} \ket{\psi_k}$, via:
\begin{align} \label{eqn:gmunu}
    \hspace{-3mm} g^{(k)}_{\mu\nu}=\Re [\mel{\psi_k}{A_\mu A_\nu}{\psi_k}{-}\mel{\psi_k}{A_\mu}{\psi_k}\mel{\psi_k}{A_\nu}{\psi_k}] .
\end{align}
Note that the AGP has a gauge freedom $A\to A+\omega$ with $\omega$ being a one-form, which enables us to always choose $A$ such that $\mel{\psi_k}{A_\mu}{\psi_k}=0$ for all $\mu$ and $k$, that is, diagonal in the Hamiltonian eigenbasis. This arises from the U(1) phase freedom for each eigenstate and thus acts diagonally in the Hamiltonian eigenbasis.

However, the resulting metric is highly sensitive to the choice of the $k$-th eigenstate. 
To characterize the eigenstate susceptibility across all eigenstates, one can average the single-eigenstate infidelity over the entire spectrum, constructing a distinct metric:
\begin{align}\label{eqn: direct sum qgt}
    \mathcal{G}_{\mu\nu}=\frac{1}{d}\sum_{k=1}^d g^{(k)}_{\mu\nu}
\end{align}
where $d$ describes the number of eigenvectors considered~\footnote{For the Ising model we will consider the full spectrum. For the latter PXP model, we will restrict ourselves to the center third of the spectrum.}; this metric was defined in~\cite{kimIntegrabilityAttractorAdiabatic2024, sharipovHilbertSpaceGeometry2026}. The resulting metric inherits the gauge invariance associated with each eigenstate, with $\omega^{(k)}{=}-\mel{\psi_k}{A_\mu}{\psi_k} dx^\mu$, and remains invariant under the local group $\text{U}^d(1)=\bigotimes_{j=1}^d \text{U}(1)$. 

Building on these insights, we elevate $\mathcal{G}_{\mu\nu}$ to a geometric object of independent standing by showing that it is (i) generated by the entanglement fidelity between two infinitesimally separated parallel transports, and (ii) equal, up to normalization, to the Killing form of $\mathfrak{gl}(d)$~\footnote{$\mathfrak{gl}(d)$ describes the Lie algebra of the general linear group of dimension $d$: GL$(d)$. Over the field of real numbers, this group describes real $d\times d$ matrices with non-zero determinant.} evaluated on the adiabatic gauge potentials. The former grants an operational meaning, while the latter fixes its residual gauge freedom intrinsically. Furthermore, a complementary perspective for understanding the emergent geometry of quantum states is based on the entanglement fidelity of the unitary transformation that induces the parallel transport.
Recall the definition of the entanglement fidelity $F_\text{ent}(U_1,U_2)=\vert{}\text{tr} (U_1^\dagger U_2)\vert{}^2/d^2$~\cite{nielsenSimpleFormulaAverage2002}.
Expanding $F_\text{ent}$ to second order between two infinitesimally close paths yields (see Supplemental Material~\cite{sm})
\begin{align}\label{eqn: Hilbert_killing metric}
    1{-}F_\text{ent}\Big(U(x,x),U(x, x{+}dx)\Big)&\approx  \frac{1}{d}\,\text{tr}A^2-\frac{1}{d^2}(\text{tr} A)^2~.
\end{align}

One may worry that the two quantities (Eqs.~\eqref{eqn: direct sum qgt},~\eqref{eqn: Hilbert_killing metric}) exhibit different gauge structures.
The entanglement fidelity $F_\text{ent}$ is invariant under a global U$(1)$ phase rather than the full U$^d(1)$ group. 
Nevertheless, one can always choose $A_\mu$ such that $\mel{\psi_k}{A_\mu}{\psi_k}=0$ for all eigenvectors. This cancels the second term in both Eqs.~\eqref{eqn:gmunu} and \eqref{eqn: Hilbert_killing metric} and ensures that $F_\text{ent}$ encodes the same metric as before. 
This is analogous to the relationship between the entanglement fidelity and the average gate fidelity~\cite{nielsenSimpleFormulaAverage2002}.

Importantly, the main feature of Eq.~\eqref{eqn: Hilbert_killing metric} is that it carries an intrinsic Lie-algebraic structure. 
We refer to $\mathcal{G}_{\mu\nu}$ as the \textit{Hilbert-Killing (HK) metric}, as it can be directly mapped to the Killing form [see Fig.\ref{fig: intro fig}(b-d)].
The Killing form $\mathcal{B}(X,Y)$ of the Lie algebra $\mathfrak{gl}(d)$ for elements $X, Y\in\mathfrak{gl}(d)$ is given by $\mathcal{B}(X,Y)\vert{}_{\mathfrak{gl}(d)}=2d\tr(XY)-2\tr(X)\tr(Y)$. Setting $X{=}A_\mu$ and $Y{=}A_\nu$, the Killing form yields $\mathcal{B}(A_\mu,A_\nu)\vert{}_{\mathfrak{gl}(d)}/(2d^2)=\mathcal{G}_{\mu\nu}$. Note that, since $\mathfrak{gl}(d)$ is not semisimple, its Killing form is degenerate: for any element $c\cdot\mathbb{1}$ in the centre, $\mathcal{B}(c\cdot\mathbb{1}, Y) = 2dc\,\mathrm{tr}\,Y - 2dc\,\mathrm{tr}\,Y = 0$ for all $Y\in\mathfrak{gl}(d)$, so the metric is insensitive to the $\mathrm{tr}\,A$ component of the AGP and, without loss of generality, the gauge choice is $\mathrm{tr}\,A = 0$. 
This explains how the HK metric has the same gauge symmetry as the entanglement fidelity. To recover the state-averaged metric, we must fix the gauge with the stronger condition that $\mel{\psi_k}{A_\mu}{\psi_k}=0$. 
Under this stronger condition, the state-averaged metric, the entanglement fidelity, and the Killing form on $\mathfrak{gl}(d)$ are fundamentally equivalent formulations of the same geometric quantity.
\begin{figure}
    \centering
    \includegraphics[width=\linewidth]{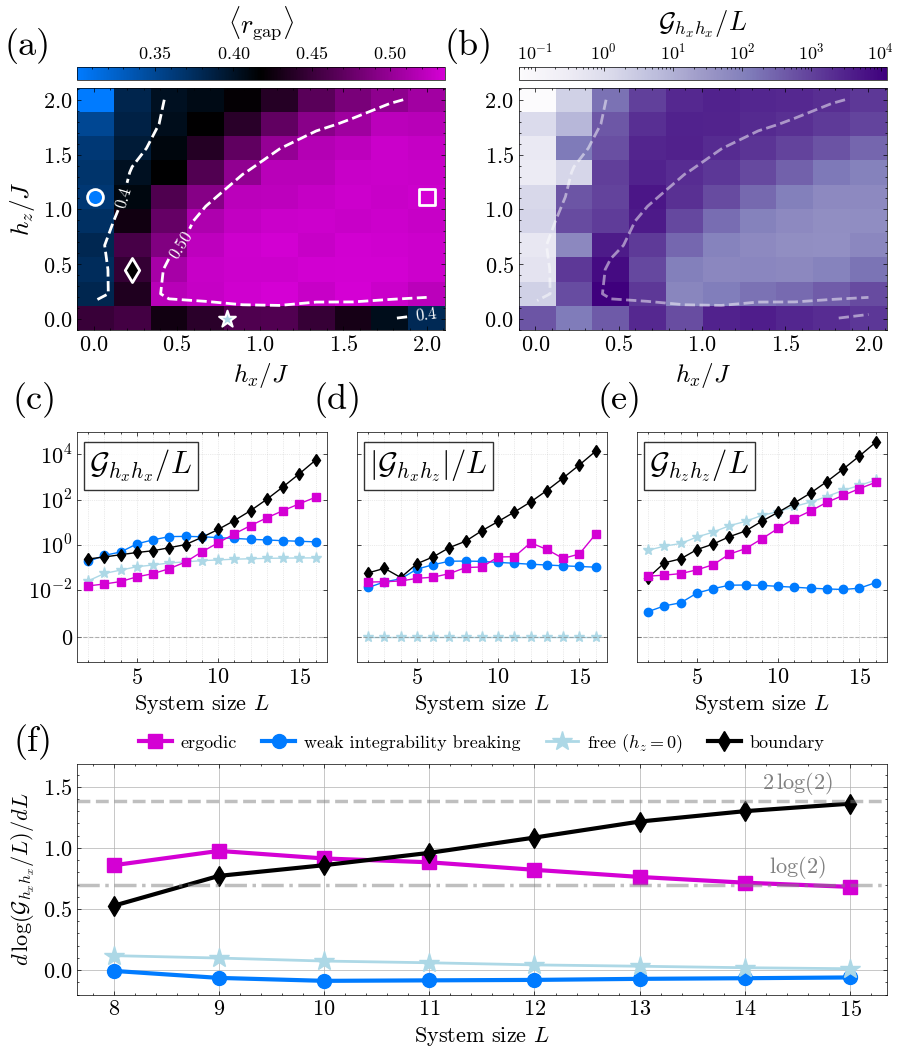}
    \caption{\textbf{Hilbert-Killing metric for the Ising spin chain.} The Ising model in Eqn.~\eqref{eqn: ising model} is parametrized by the fields $(h_x/J, h_z/J)$. In \textbf{(a,b)} we compute the mean gap ratio $\expval{r_\text{gap}}$ and the HK component $\mathcal{G}_{h_xh_x}$ for $L=15$ as a function of the parameters $(h_x/J, h_z/J)$, where $h_{x,z}/J\in [0.01,2]$. We identify that the biggest response of the HK metric is given precisely at the phase boundary between the integrable and ergodic limits. 
    Given the multi-directionality of the HK metric, we can also see how different directions of the perturbations act on the averaged eigenstate deformations (see panels \textbf{(c-e)}, plotted on a symlog scale). For instance, while the $\mathcal{G}_{h_xh_x}$ component \textbf{(c)} demonstrates clear distinction between all three regions, the exponential separation between the ergodic and integrable phases is not reproduced in \textbf{(d,e)}. Additionally, the off-diagonal component \textbf{(d)} also illustrates that the scaling of the ergodic regime is suppressed, while in the free regime, by parity symmetry, it is equal to zero for all system size scales.
    In \textbf{(f)}, we choose four points corresponding to the integrable (blue circle), free (light-blue star, $h_z=0$), ergodic (purple square), and the boundary region (black diamond) and study the scaling behaviour by plotting the logarithmic derivative of $\mathcal{G}_{h_xh_x}$, showcasing the fastest growth scaling at the crossover regime. The factors of $\log(2)$ (dashed-dotted) and $2\log(2)$ (dashed) gray lines denote the expected asymptotic values for the growth of HK metric components being ergodic and maximal, respectively~\cite{leblondUniversalityOnsetQuantum2021}. The above results also hold for all components~\cite{sm}. The precise values of the points can be found in~\cite{sm}.   
    }
    \label{fig: ising}
\end{figure}

When the Hamiltonian has a single parameter, the HK metric exactly reproduces the results of the norm of the AGP as $\tr A^2{=}\norm{A}^2$~\cite{pandeyAdiabaticEigenstateDeformations2020}.
The key advantage of the HK metric is that it provides a more fine-grained probe of the eigenstate susceptibility.
In particular, for $n$ parameters, the HK metric will contain $\mathcal{O}(n^2)$ components, which allows a more extensive study of the ergodic-integrable transition as well as detect the interplay of different perturbations.
For example, the off-diagonal terms of the HK metric provide a lower bound on the product of variances under different adiabatic deformations.
More concretely, let us define the spectrum-averaged variance of the AGP $A_\mu$ as $\sigma_{A_\mu}^2=d^{-1}\tr ( \left[A_\mu-\expval{A_\mu}\right]^2)$ with $\expval{A_\mu}=d^{-1}\tr A_\mu$.
This variance quantifies how much eigenstate deformations fluctuate around their mean value.
By the Cauchy-Schwarz inequality, we are guaranteed that~\cite{sm}:
\begin{align}
    \label{eqn: geometric uncertainty}\sigma_{A_\mu}^2\sigma_{A_\nu}^2\geq \abs{\mathcal{G}_{\mu\nu}}^2.
\end{align}
Equality of the bound is provided for the diagonal components of the Hilbert-Killing metric, where $\sigma_{A_\mu}^2= \abs{\mathcal{G}_{\mu\mu}}$.
This inequality shows that the HK metric quantifies the total overall amount of eigenstate variation along different parameter perturbations.
In other words, the diagonal HK components are the spectrum-averaged variances of the AGP, and the off-diagonals measure how those deformations are correlated.
This provides an intuition for why its behavior captures the ergodic/integrable behavior of a system. 
In ergodic systems, small changes to the parameters strongly modify the eigenvectors, so the variance is large, whereas in integrable systems, the existence of locally conserved quantities suppresses such changes. 
The above Cauchy-Schwarz inequality also allows us to directly provide a bound on the system size scaling on every component of the HK metric~\cite{sm}
\begin{align}
    \label{eqn: extended bound}
    |\mathcal{G}_{\mu\nu}|\leq \frac{\Vert\partial_\mu\hat H\Vert\,\Vert\partial_\nu\hat H\Vert}{4\mu_c^2},
\end{align}
where $\mu_c$ is a numerical cutoff in the energy resolution, chosen as $L/d$, similar to~\cite{pandeyAdiabaticEigenstateDeformations2020}. For the models considered, we can compute this bound analytically~\cite{sm}.
As we will see below, the HK identifies these regimes through its specific scaling behaviour.

\textit{Numerical investigation of the HK metric--- }
We now demonstrate the HK metric as a probe of the onset of ergodicity in many-body quantum systems.
We begin by considering the spin-1/2 one-dimensional mixed-field Ising model
\begin{align}
    \label{eqn: ising model}
    \hat{H} =  J\sum_{j=1}^{L-1} \sigma^z_{j+1}\sigma^z_{j}+h_z\sum_{j=1}^{L}\sigma^z_j+h_x\sum_{j=1}^{L}\sigma^x_j.
\end{align}
where $\sigma^\alpha_i$ is the $\alpha$ Pauli operator on site $i$, $J$ is the Ising interaction strength, and $h_{x}$ and $h_z$ are the transverse and longitudinal fields, respectively. 
This Hamiltonian is free whenever $h_z=0$, diagonal in $\sigma^z$ when $h_x=0$, and hosts a robust ergodic phase when both are large~\cite{kimBallisticSpreadingEntanglement2013}; exemplary points of the free and ergodic regimes are denoted by the light-blue star and the purple square in Fig.~\ref{fig: ising}(a), respectively.
The integrable (including free) and ergodic regimes can be identified by the gap ratio approaching the Poissonian~\cite{berryLevelClusteringRegular1977} and Gaussian~\cite{bohigasCharacterizationChaoticQuantum1984a} random matrix ensemble averages~\cite{atasDistributionRatioConsecutive2013}. 

Exploring this phase diagram with the HK metric and mean gap ratio [see Fig.~\ref{fig: ising}(a,b)], we observe that the HK metric accurately captures the boundary between the ergodic and integrable regimes. 
This boundary manifests itself as a much larger value of the metric.
Note that, while we focus on one particular element $\mathcal{G}_{h_xh_x}$ of the metric in Fig.~\ref{fig: ising} for the phase diagram, our observations apply across all other elements~\cite{sm}.
This is consistent with the fidelity-susceptibility picture of the integrability-breaking crossover developed for both classical and quantum systems in~\cite{pandeyAdiabaticEigenstateDeformations2020,kimDefiningClassicalQuantum2026}.

\begin{figure}
    \centering
    \includegraphics[width=\linewidth]{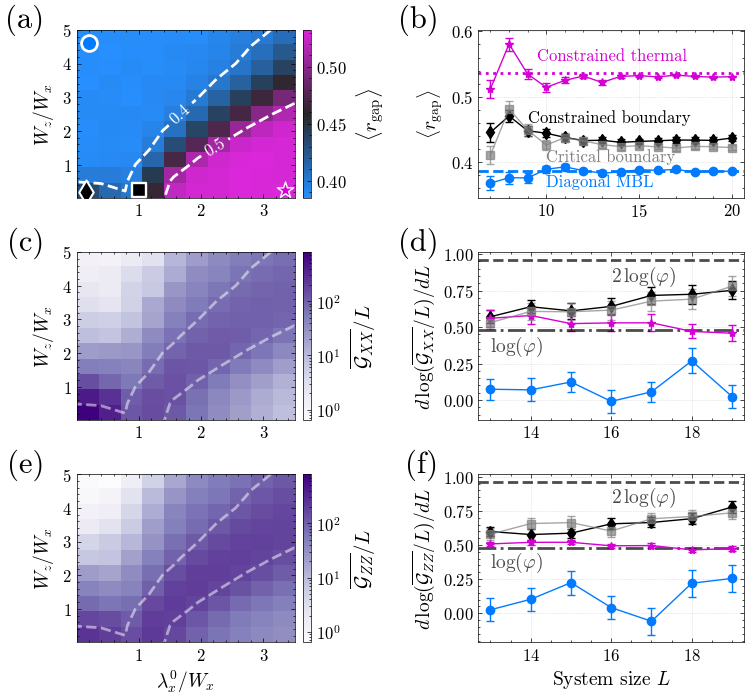}
    \caption{\textbf{Hilbert-Killing metric for locally constrained systems.} We analyze the phase diagrams $(\lambda_x^0/W_x, W_z/W_x)$ \textbf{(a,c,e)} for fixed system size $L=17$ and the corresponding scaling behavior \textbf{(b,d,f)} for the PXP model in Eqn.~\eqref{eqn: pxp}. For shortened notation we defined $X_j=\hat{P}\sigma^x_j\hat{P}$ and $Z_j=\hat{P}\sigma^z_j\hat{P}\equiv \sigma^z_j$. To prevent outliers from strongly affecting the HK metric, we focus on the median HK metric denoted by the overline as $\overline{\mathcal{G}_{\mu\nu}}$ alongside the median absolute deviation. Similar to the Ising model analysis, we find that the largest (median) HK response occurs between the integrable and ergodic phases. Studying the scaling behaviour for the constrained thermal (purple stars), constrained boundary (black diamonds), critical boundary (black transparent squares), and diagonal MBL (blue circles), we observe a robust scaling signature at the ergodic-integrable boundary. For each point in the phase diagram and also for all system sizes in the scaling analysis, we average over 100 disorder realizations and only take the center third of the many-body spectrum, akin to~\cite{chenHowDoesLocally2018}. Similar to Fig.~\ref{fig: ising}, we also include the expected upper bounds on the scaling, which are proportional to $\log(\varphi)$ and $2\log(\varphi)$, respectively, with $\varphi\approx 1.618$, the golden ratio. The precise values of the points can be found in the supplemental materials~\cite{sm}.}
    \label{fig: pxp}
\end{figure}
This large response suggests that the scaling of the HK metric with system size might exhibit different behaviors for different (non-)thermalizing regimes.
This has been observed for the AGP in free, integrable, and ergodic systems \cite{pandeyAdiabaticEigenstateDeformations2020}, motivating an investigation of the full HK metric.
Indeed, all components of the HK metric show markedly distinct responses [see Fig.~\ref{fig: ising}(c-e)]. For instance, the vanishing of the off-diagonal component $\mathcal{G}_{h_x h_z}$ at the free point [see Fig.~\ref{fig: ising}(d)] follows from the $\mathbb{Z}_2$ spin-flip symmetry of the free model: since $\partial\hat{H}/\partial h_x = \sum_j \sigma^x_j$ is even and $\partial\hat{H}/\partial h_z = \sum_j \sigma^z_j$ is odd under this symmetry, the product $A_{h_x}A_{h_z}$ is block off-diagonal between parity sectors, causing all diagonal matrix elements to vanish by a parity selection rule. Conversely, $\mathcal{G}_{h_z h_z}$ [see Fig.~\ref{fig: ising}(e)] is strongly enhanced at the free point because $h_z$ is the integrability-breaking direction of the free model, rendering the eigenstates maximally susceptible to changes along it.

Finally, it is important to note that the scaling distinction of the AGP was initially studied for integrability-preserving perturbations in the Ising model~\cite{pandeyAdiabaticEigenstateDeformations2020}.
Indeed, while we observe the same scaling separation when probing $\mathcal{G}_{h_xh_x}$, this distinction disappears when focusing on other combinations of perturbations, namely those encoded in $\mathcal{G}_{h_xh_z}$ and $\mathcal{G}_{h_zh_z}$. 
Notably, the behaviour of fastest growth scaling is consistent across all components.
To quantify this relationship, we consider the logarithmic derivative of the HK metric $\mathcal{G}_{h_xh_x}$ for each regime [see Fig.~\ref{fig: ising}(f)].
Whereas the ergodic regime approaches the predicted value of $\log(2)$~\cite{pandeyAdiabaticEigenstateDeformations2020}, and the integrable regime has a near-zero growth rate, the HK metric at the ergodic-integrable boundary increases super-exponentially with system size after reaching the value of $2\log(2)$ in agreement with the aforementioned extended bound in Eq.~\eqref{eqn: extended bound}. Note that the derived bound caps the asymptotic growth rate of the HK metric; hence, at finite $L$ the local slope can temporarily exceed this value~\cite{sm}.
This demonstrates the power of the HK metric, in that it encodes the response of the system to multiple perturbations, enabling a more comprehensive understanding of the components of the Hamiltonian that drive thermalization.

To explore the extent to which the studied features of the HK metric are universal, we now turn to a constrained PXP model which can host a variety of integrable many-body phases upon the introduction of disorder
~\cite{chenHowDoesLocally2018, turnerWeakErgodicityBreaking2018}
\begin{align}
    \label{eqn: pxp}
    \hat{H}_\text{PXP}&=\sum_{j=1}^{L}(\lambda_x)_j \hat{P}\sigma^x_j\hat{P}+\sum_{j=1}^{L}(\lambda_z)_j\sigma^z_j
\end{align}
Here the coefficients $(\lambda_x)_j, (\lambda_z)_j$ are drawn uniformly from $\lambda_x\in \lambda_x^0{+}[-W_x, W_x]$ and $\lambda_z\in [-W_z,W_z]$, respectively. 
The projector $\hat{P}=\prod_j[1- (1-\sigma_{j}^z)(1-\sigma_{j+1}^z)/4]$ constrains neighboring spins such that not both can be in the $\sigma^z=-1$ state, reducing the Hilbert space dimension for open boundary conditions to $d{=}F_{L+2}{<}2^L$, where $F_{L}$ is the $L$-th Fibonacci number and $F_L\sim\varphi^L$ with $\varphi\approx 1.618$ being the golden ratio~\cite{chenHowDoesLocally2018}.

The $(\lambda_x^0/W_x, W_z/W_x)$ phase diagram exhibits three different regimes: a diagonal MBL, a constrained MBL, and the ergodic regime~\cite{chenHowDoesLocally2018}.
Besides distinguishing between the localized and ergodic regimes, the mean gap ratio identifies two regions consistent with the onset of ergodicity [see Fig.~\ref{fig: pxp}(a,b)]:
one between the localized and ergodic regimes (black square, denoted as critical boundary), and the other near $W_z=\lambda^0_x=0$ (black diamond, denoted as constrained boundary).



Crucially, the observations in the PXP model mirror those of the Ising model studied above: the (median) HK metric, denoted as $\overline{\mathcal{G}_{\mu\nu}}$, has the strongest signatures at the integrable-ergodic boundary~[see Fig.~\ref{fig: pxp}(c,e)]. 
This observation holds both along the critical boundary as well as the constrained boundary. 
To understand the asymptotic behavior of the metric across different physical regimes, we again perform a finite-size scaling analysis~[see Fig.~\ref{fig: pxp}(d,f)]. 
Deep within the ergodic phase, the HK metric exhibits a characteristic exponential scaling with system size. 
At the ergodic-integrable phase boundary, however, the metric displays the transient super-exponential system-size scaling. Within that regime, the HK metric, however, has not yet reached the derived value of $2\log(\varphi)$. 

On the integrable side, the scaling features are more subtle. 
Here, we observe a distinctly slower growth rate accompanied by pronounced finite-size fluctuations, which preclude a sharp numerical determination of a single scaling form. 
We note that this behavior is consistent with the presence of rare outlier events~\cite{sm}. 
When coupled with the exponential sensitivity of the HK metric, these outliers can introduce significant variance into the disorder average.
Unlike the mean gap ratio, the HK metric shows the strongest system size scaling specifically pinned to the ergodic-integrable boundary~[see Fig.~\ref{fig: pxp}(d,f)]. 
This matches our numerical findings for the Ising model, suggesting that this scaling signature is not specific to the Ising chain.

\textit{Conclusion --- }In this work, we have shown that quantum many-body phases are imprinted on an emergent metric structure via the Hilbert-Killing metric. 
By going beyond the norm of a single adiabatic gauge potential to the full Hilbert-Killing metric, whose off-diagonal components capture correlations between eigenstate deformations along different parameter directions, we obtain a robust and interpretable tool for characterizing the onset of ergodicity.
In particular, we observe a transient super-exponential scaling at the ergodic-integrable boundary regime for two distinct models: the Ising and PXP models.

In conclusion, our results suggest that the quantum geometry, endowed with the HK metric, provides a natural bridge between ergodicity, information theory, and differential geometry.

\textit{Acknowledgements --- }We thank all members of the Rimbach-Russ, Bosco, Veldhorst, and Vandersypen groups for providing valuable insights. 

\textit{Funding --- }M.R.-R., S.B., E.V., and C.V.M. acknowledge that the EU partly supported this research through the Horizon Europe QLSI2 project and was partly sponsored by the Army Research Office under Award Number: W911NF-23-1-0110. M.R.-R. and E.V. acknowledge support from the Dutch Research Council (NWO) under Award Number Vidi TTW 22204. The views and conclusions contained in this document are those of the authors and should not be interpreted as representing the official policies, either expressed or implied, of the Army Research Office or the U.S. Government. The U.S. Government is authorized to reproduce and distribute reprints for Government purposes, notwithstanding any copyright notation herein. F.~M.~  acknowledges support from the Netherlands Organisation for Scien-
tific Research (NWO/OCW), as part of Quantum Limits
(project number SUMMIT.1.1016).

\textit{Author contributions --- }C.V.M. and E.V. performed the theoretical computations and numerical simulations with inputs from F.M., M.R.-R., and S.B. C.V.M. conceived and F.M. and M.R.-R. supervised the project. C.V.M. wrote the manuscript with input from all the authors.

\textit{Data availability ---}The data are publicly available at \footnote{Online repository: \url{https://doi.org/10.5281/zenodo.21822427}}.

\appendix 

\newpage

\begin{widetext}
    \begin{center}
\textbf{\large Supplemental Material: Geometric signatures of the onset of many-body ergodicity}
\end{center}
\end{widetext}

\tableofcontents

\section{Differential geometry and HK metric}
\label{app:1 diff geo and hk}

\subsection{Derivation HK metric}
\label{app: HK derivation }
Geometric approaches, where dynamics resulting from a parameter
dependent Hamiltonian are mapped to the evolution of the parameters themselves, provide a powerful framework for characterizing and informing quantum dynamics~\cite{ventura-meinersenQuantumGeometricProtocols2025, meinersenUnifyingAdiabaticStatetransfer2025, ventura-meinersenMultilevelSpectralNavigation2026a}. Here, we will show the derivation of the Hilbert-Killing metric, for which we need the expression of the translation operator in parameter space
\begin{align}
    U(x_\text{i},x_\text{f})=\mathcal{P}\exp\left(-i\int_{x_\text{i}}^{x_\text{f}} A\right).
\end{align}
We aim to compare two parallel transports and check the distance between them, i.e., we aim to study $F_\text{ent}=|\tr(U^\dagger(x,y)U(x,z))|^2/d^2$, where we expand $z\approx y +dy$, which yields
\begin{align}
    U^\dagger(x,y)U(x,y+dy)\approx U^\dagger(x,y)U(y,y+dy)U(x,y), 
\end{align}
where, by the use of the trace, we find that the entanglement fidelity reduces to
\begin{align}
    F_\text{ent}\approx|\tr(U(x,x+dx))|^2/d^2,
\end{align}
where we have relabeled $y\to x$. If we perform a second-order expansion around the starting point, we find
\begin{align}
    U(x,x{+}dx)&\approx\exp\left(-i\left[A_\mu dx^\mu +\frac{1}{2}(\partial_\alpha A_\mu)dx^\mu dx^\alpha\right]\right)\\
    &\equiv \exp\left(-i\left[A +\frac{1}{2}\partial A\right]\right)\\
    &\approx \mathbb{1}_d-i\left[A +\frac{1}{2}\partial A\right]+\frac{1}{2}\left(-i\left[A +\frac{1}{2}\partial A\right]\right)^2 \\
    &\approx \left(\mathbb{1}_d-\frac{1}{2}A^2\right)-i\left(A+\frac{1}{2}\partial A\right).
\end{align}
Taking the trace, we find a complex number whose absolute value squared is given by
\begin{align}
    \abs{\tr U(x,x+dx)}^2&\approx  \left(d-\frac{1}{2}\tr A^2\right)^2+\left(\tr A+\frac{1}{2}\tr \partial A\right)^2\\
    &\approx \left(d^2-d\tr A^2\right)+\left(\tr A\right)^2\\
    &= d^2 - \Big(d\tr A^2 - (\tr A)^2\Big).
\end{align}
Note that $\tr \partial A=\partial \tr A$, with $\tr A \in \mathbb{R}$, is real-valued. We find that the term $\propto \tr \partial A$ does not contribute at second order. For the entanglement fidelity, we find that
\begin{align}
    F_\text{ent}&\approx\frac{|\text{tr}(U(x,x+dx))|^2}{d^2}\\
    &\approx \frac{d^2 - \Big(d\tr A^2 - (\tr A)^2\Big)}{d^2}\\
    &=1-\frac{1}{d^2}\Big(d\tr A^2 - (\tr A)^2\Big).
\end{align}
Using the Hilbert-Schmidt inner product $\expval{M,N}_\text{HS}{=}\tr(M^\dagger N)$, we find
\begin{align}
    &\mathcal{G}_{\mu \nu}dx^\mu dx^\nu = \frac{1}{d^2}\Big(d\,\text{tr} A^2-(\text{tr} A)^2\Big)\\
    &\equiv\frac{1}{d^2}\left(\expval{\mathbb{1}_d,\mathbb{1}_d}_\text{HS}\,\expval{A,A}_\text{HS}-\Big(\expval{A,\mathbb{1}_d}_\text{HS}\Big)^2\right).
\end{align}
We note that given the Cauchy-Schwarz inequality
\begin{align}
    \Big(\expval{A,\mathbb{1}_d}_\text{HS}\Big)^2\leq \expval{\mathbb{1}_d,\mathbb{1}_d}_\text{HS}\,\expval{A,A}_\text{HS},
\end{align}
we find that the HK metric is always positive semi-definite $\mathcal{G}\geq0$ as expected from a distance metric. Using the gauge invariance of shifting the adiabatic gauge potential $A\to A+\omega$, we can set $\tr A=0$, and hence we find that $\mathcal{G}_{\mu\nu}dx^\mu dx^\nu=(\tr A^2)/d$, which trivially showcases the positivity of the HK metric. The Hilbert-Killing metric is hence the expectation value of the (squared) adiabatic gauge potential in terms of the state $\tilde{\rho}=\sum_n \ketbra{\psi_n}/d$ as $\mathcal{G}_{\mu\nu}dx^\mu dx^\nu=\tr(\tilde{\rho} A^2)$. We will motivate the origin of this state below.

\subsection{Bound of HK metric}
\label{app: geo uncertainty}
\subsubsection{Average density matrices}

To address what the physical consequences of the HK metric are, we investigate the space of density matrices, which are described by the underlying parameters $x=x^\mu$ as $\rho\equiv \rho(x)$. We will derive a geometric correlation bound of the HK metric, which we interpret as an uncertainty principle similar to the Robertson-Schrödinger relation of quantum observables. 

For that, we need to introduce the relevant space of states. A quantum mechanical state $\rho$ is an element of $\text{D}(\mathcal{H})$, which is defined as follows
\begin{align}
    \text{D}(\mathcal{H})=\{\rho\in \mathcal{H}| \tr \rho = 1, \rho \geq0, \rho^\dagger=\rho\}.
\end{align}
We want to make a further distinction, as states $\rho$ can have different ranks. For instance, pure states are rank-1 states. Hence, we define the space $\mathcal{D}_{k,d}$ as follows~\cite{houLocalGeometryQuantum2024}
\begin{align}
    \mathcal{D}_{k,d}(\mathcal{H})=\{\rho\in \text{D}(\mathcal{H})|\rank \rho =k\},
\end{align}
where $d=\dim \mathcal{H}$ is the dimension of the underlying Hilbert space. For our analyses, we restrict ourselves to the full-rank case, ie, $\mathcal{D}_{d,d}(\mathcal{H})\equiv \mathcal{D}$. We wish to study a statistical ensemble of density matrices in this space and compute its mean
\begin{align}
    \Bar{\rho}=\int_\mathcal{D}d\rho \,\mu(\rho) \,\rho,
\end{align}
where $\mu(\rho)=1/\text{vol}(\mathcal{D})$ constitutes the (constant) integration measure resulting from Jaynes' maximization of the Shannon entropy $-\int_\mathcal{D}\mu(\rho)\log \mu(\rho)$ with no additional constraints besides the normalization of the density matrix~\cite{jaynesInformationTheoryStatistical1957,jaynesInformationTheoryStatistical1957a}. Using the maximization strategy, one may also envision adding further constraints by, for instance, fixing the total energy through a Lagrange multiplier. By diagonalizing the density matrix $\rho=U\Lambda U^\dagger$ we can further incorporate the positivity and hermiticity constraint, yielding an average density matrix
\begin{align}
    \Bar{\rho}=\frac{1}{\text{vol}(\mathcal{D})}\int d\Lambda \,\theta(\Lambda)\delta(1-\tr \Lambda)\int_{U(d)}dU\,U\Lambda U^\dagger, 
\end{align}
where 
\begin{align}
    \text{vol}(\mathcal{D})&=\iint dU\,d\Lambda\; \theta(\Lambda)\delta(1-\tr \Lambda)\\
    &=\int d\Lambda\; \theta(\Lambda)\delta(1-\tr \Lambda)    
\end{align}
is the volume element and $\int dU=1$, due to normalization. $\theta(x)$ represents the Heaviside theta function and enforces positivity of the eigenvalues. We can make use of the twirling over the Haar measure $dU$ to get
\begin{align}
    \Bar{\rho}&=\frac{1}{\text{vol}(\mathcal{D})}\int d\Lambda \,\theta(\Lambda)\delta(1-\tr \Lambda)\left(\frac{\tr \Lambda}{d}\mathbb{1}\right)=\frac{1}{d}\mathbb{1}.
\end{align}
In the many-body ergodic regime, we assume that $\rho\approx \Bar{\rho}$, which can be decomposed into the eigenvectors of the Hamiltonian to get the state
\begin{align}
    \tilde{\rho}= \Bar{\rho}=\frac{1}{d}\sum_{n=1}^d\ketbra{\psi_n}.
\end{align}
We note that $\tilde{\rho} = \mathbb{1}/d$ is the infinite-temperature (maximally mixed) state; identifying it with the ergodic steady state is an assumption of equal spectral weight across all eigenstates, not a direct consequence of the eigenstate thermalisation hypothesis, which at finite energy density predicts a microcanonical rather than infinite-temperature ensemble. In the numerical results, we therefore restrict to a microcanonical energy window (at infinite temperature), which for the Ising model refers to the entire spectrum and for the PXP model the center third to be aligned with the results of~\cite{chenHowDoesLocally2018}.
\begin{figure*}
    \centering
    \includegraphics[width=\textwidth]{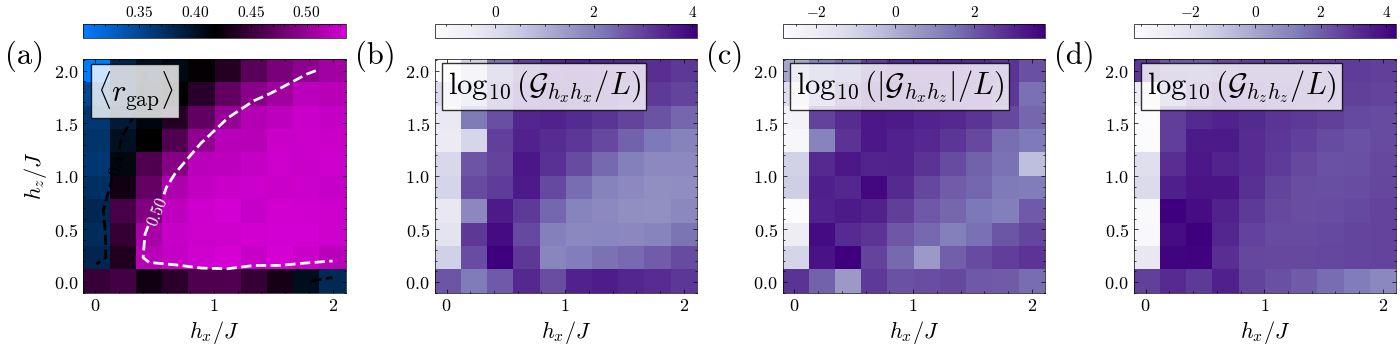}
    \caption{The HK metric captures average eigenstate sensitivities with respect to all parameters in the Hamiltonian for $L=15$ as in the main text. \textbf{(a-d)} We study the phase space $(h_x/J, h_z/J)$ of the mixed-field Ising model in terms of the mean gap ratio and the HK metric components.}
    \label{fig: ising supp}
\end{figure*}
\begin{figure*}
    \centering
    \includegraphics[width=\textwidth]{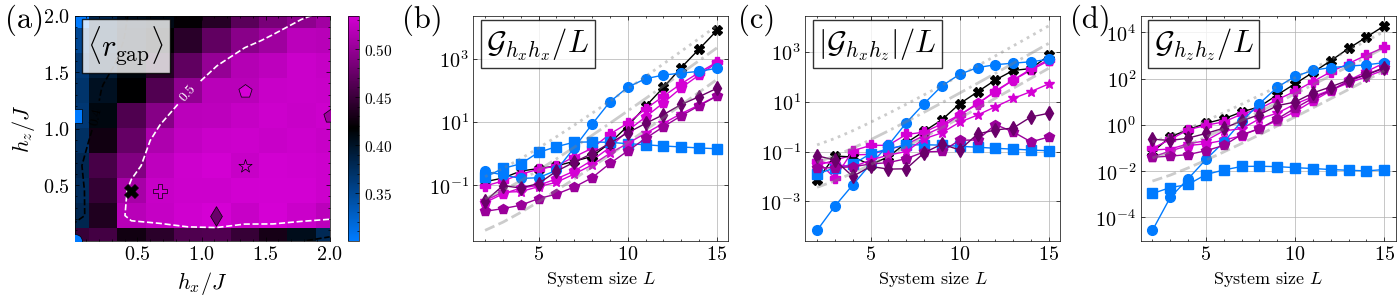}
    \caption{Further numerical analysis of Hilbert-Killing response for the mixed-field Ising model. At the integrable points, the HK metric components plateau in their growth. As in the main text, for points in the crossover regime we find the largest response as the system size $L$ increases.}
    \label{fig: ising more points}
\end{figure*}
\begin{figure*}
    \centering
    \includegraphics[width=\linewidth]{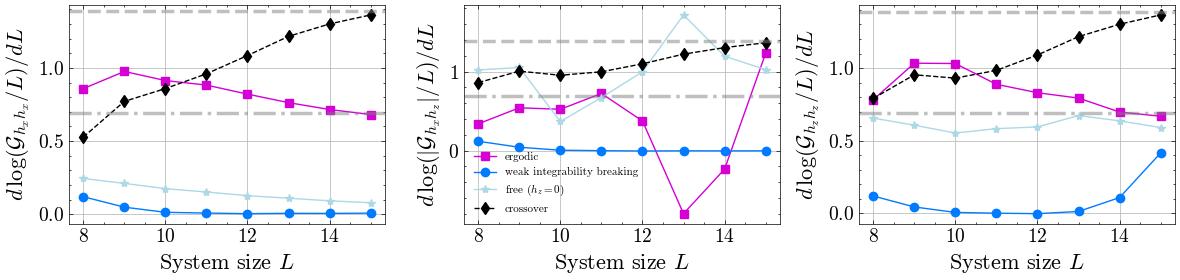}
    \caption{Logarithmic derivative of the HK metric components for the mixed-field Ising model. The crossover point shows the strongest scaling in all components at the largest sizes accessible, with the off-diagonal component behaving non-monotonically in the ergodic and free regimes. The non-monotonic behaviour of the free regime can be understood as follows: $|\mathcal{G}_{h_x h_z}|$ passes through zero in this region, which is exactly zero at the free point, by the parity argument in the main text, but is not here due to numerical precision, so $d\log |\mathcal{G}|/dL$ is ill-conditioned near the sign change, and the outliers are numerical rather than physical. Nevertheless, the boundary scaling is robust.}
    \label{fig: ising supp scaling}
\end{figure*}
\subsubsection{Geometric uncertainty principle}
We can define a set of bounded operators $\{A_\mu\}$ as linear operators acting on states $\rho \in \mathcal{D}_{d,d}$. We will interpret the set of bounded operators as generators of translations in the tangent space $\text{T}\mathcal{D}_{d,d}$ of the full-rank states, where 
\begin{align}
    \text{T}\mathcal{D}_{d,d}=\bigcup_x \text{T}_x\mathcal{D}_{d,d}.
\end{align}
The space of quantum states is known to carry both a Riemannian and a symplectic structure~\cite{ciagliaDifferentialGeometryQuantum2019, heydariGeometricFormulationQuantum2016, haegemanGeometryMatrixProduct2014, houLocalGeometryQuantum2024}. This leads us to investigate the variance of the tangent generators $\sigma_{A_\mu}^2$ as an observable for the sensitivity of chaotic fluctuations. For general full-rank states, we find
\begin{align}
    \sigma_{A_\mu}^2&= \expval{\left(A_\mu-\expval{A_\mu}\right)^2}\\
    &= \tr (\rho \left[A_\mu-\expval{A_\mu}\right]^2)\\
    &= \expval{X_\mu,X_\mu}_\text{HS},
\end{align}
where we defined $X_\mu:=\sqrt{\rho}\left(A_\mu-\expval{A_\mu}\right)$ under the Hilbert-Schmidt inner product $\expval{A,B}_\text{HS}=\tr(A^\dagger B)$. To obtain a lower bound, we can apply the Cauchy-Schwarz inequality
\begin{align}
    \sigma_{A_\mu}^2\sigma_{A_\nu}^2&=\expval{X_\mu,X_\mu}_\text{HS}\expval{X_\nu,X_\nu}_\text{HS}\\
    &\geq |\expval{X_\mu,X_\nu}_\text{HS}|^2.
\end{align}
We can compute the overlap as follows (we define $\delta A_\mu := A_\mu-\langle A_\mu\rangle$)
\begin{align}
    \expval{X_\mu,X_\nu}_\text{HS} &= \tr(X^\dagger_\mu X_\nu)\\
    &= \tr(\rho \,\delta A_\mu\, \delta A_\nu) \\
    &= \frac{1}{2}\tr(\rho\{\delta A_\mu, \delta A_\nu\})+\frac{1}{2}\tr(\rho [\delta A_\mu, \delta A_\nu])\\
    &=\frac{1}{2}\expval{\{\delta A_\mu, \delta A_\nu\}}+\frac{1}{2}\expval{[\delta A_\mu, \delta A_\nu]}
\end{align}
For $\rho\propto \mathbb{1}$, the commutator vanishes due to cyclicity of the trace and we find
\begin{align}
    |\expval{X_\mu,X_\nu}_\text{HS}|^2&= \abs{\frac{1}{2}\expval{\{\delta A_\mu, \delta A_\nu\}}}^2\\
    &=\abs{\expval{A_\mu A_\nu}-\expval{A_\mu}\expval{A_\nu}}^2,
\end{align}
where we find that the product of variances is lower bounded by the HK metric
\begin{align}
    \sigma_{A_\mu}^2\sigma_{A_\nu}^2\geq \abs{\mathcal{G}_{\mu\nu}}^2.
\end{align}
Equality of the bound is provided for the diagonal components of the Hilbert-Killing metric, where $\sigma_{A_\mu}^2= \abs{\mathcal{G}_{\mu\mu}}$.

\subsection{Bound on the Hilbert-Killing metric}
\label{sec:hkbound}

The norm of the regularised adiabatic gauge potential cannot grow arbitrarily fast with system size. Writing~\cite{pandeyAdiabaticEigenstateDeformations2020} $\Vert A_\mu\Vert^2=\int\mathrm{d}\omega\, \omega^2(\omega^2+\mu_c^2)^{-2}|f_\mu(\omega)|^2$ with the cutoff $\mu_c$ and using $x^2/(x^2+1)^2\le 1/4$, one finds $\Vert A_\mu\Vert^2\le\Vert\partial_\mu\hat{H}  \Vert^2/(4\mu_c^2)$. Since the diagonal components of the Hilbert-Killing metric coincide with the AGP norm, this immediately gives $\mathcal{G}_{\mu\mu}\le \Vert\partial_\mu\hat H\Vert^2/(4\mu_c^2)$. The off-diagonal components follow from positive semi-definiteness of the metric, $|\mathcal{G}_{\mu\nu}|^2\le\mathcal{G}_{\mu\mu}\mathcal{G}_{\nu\nu}$, yielding the generalised bound
\begin{equation}
  |\mathcal{G}_{\mu\nu}|\;\le\;
  \frac{\Vert\partial_\mu\hat H\Vert\,\Vert\partial_\nu\hat H\Vert}
  {4\mu_c^2}.
  \label{eq:hkbound}
\end{equation}
For the Ising chain, one has $\Vert\partial_{h_x}\hat H\Vert^2=\Vert\partial_{h_z}\hat H\Vert^2=L$, so that with $\mu_c=L/d$ and $d=2^L$ all three independent components obey
\begin{equation}
  |\mathcal{G}_{h_ih_j}|
  \leq \frac{L}{4\mu_c^2}=\frac{4^{L}}{4L},
  \label{eq: isingbound}
\end{equation}
and correspondingly we might directly bound the derivative $\mathrm{d}\log|\mathcal{G}_{\mu\nu}|/\mathrm{d}L\le 2\log2-1/L$. 

However, for any functions $f,g: f\leq g$ it is not guaranteed that their derivatives follow $f'\leq g'$. Hence we need to be careful about how, within finite difference methods, the slope is bounded as well. For the purposes of this we will choose $f(L):=|\mathcal{G}|/L$ and $g(L):=4^L/(4L^2)$, which then are related as $f\leq g$ given the bound above. The same holds true for $\log f\leq \log g$ as the logarithm is monotonically increasing. We will proceed to define further functions:
\begin{align}
    \Delta(L) &= \log \left[\frac{g(L)}{f(L)}\right] \\
    s_\gamma(L) &= \dv[]{\log \gamma(L)}{L} \\
    \dv[]{\Delta}{L}&=s_g(L) - s_f(L),
\end{align}
where $\Delta(L)$ defines the gap that turns the bound into an equality and $s_\gamma$ represents the slope of the functions $\gamma=f,g$. We can now provide the finite difference update (here: forward derivative) in terms of these functions:
\begin{align}
    \Delta(L+1)&=\Delta(L) + \dv[]{\Delta}{L} \\
    &= \Delta(L) + [s_g(L)-s_f(L)] \\
    &\equiv \Delta(L) + e(L).
\end{align}
Repeating this step results in $\Delta(L)=\Delta(L_0)+\sum_{\ell=L_0}^{L-1} e(\ell)$ for any starting point $L_0$. Due to the inequality, the gap is always positive-semidefinite and hence
\begin{align}
    0\leq \Delta(L) &= \Delta(L_0) + \sum_\ell e(L)\\
    \Rightarrow -\Delta(L_0) &\leq \sum_\ell e(L_0).
\end{align}
Reshuffling this inequality yields
\begin{align}
    \sum_\ell s_f \leq \sum_\ell s_g + \Delta(L_0).
\end{align}
To pose an asymptotic statement, we may average over $N=L-L_0$ and take $N\to \infty$, which yields
\begin{align}
    \lim_{N\to \infty} \left[\frac{1}{N}\sum_\ell s_f\right] \leq \lim_{N\to \infty} \left[\frac{1}{N}\sum_\ell s_g\right] \leq 2\log 2,
\end{align}
where the component $\lim_{N\to \infty} [\Delta(L_0)/N]=0$ as $\Delta(L_0)$ is $N$-independent. This means that in the long-run average, the slope of the $f$, ie $s_f$, is at most $2\log 2$.

The ergodic regime saturates only the milder scaling~\cite{pandeyAdiabaticEigenstateDeformations2020}: $\mathcal{G}_{\mu\nu}\propto\mu_c^{-1}\propto d/L$, i.e.\ $\log 2$, so that the enhanced growth observed in the boundary region remains within, and is capped by, Eq.~\eqref{eq: isingbound}. For the PXP model, the argumentation is similar, but now, as the dimension scales with the golden ratio $\dim \mathcal{D}\sim \varphi^L$, we find that the scaling factors are proportional to $\log(\varphi)$.

\section{Details on numerical analysis}
\subsection{Additional numerical results}
\begin{figure}
    \centering
    \includegraphics[width=\linewidth]{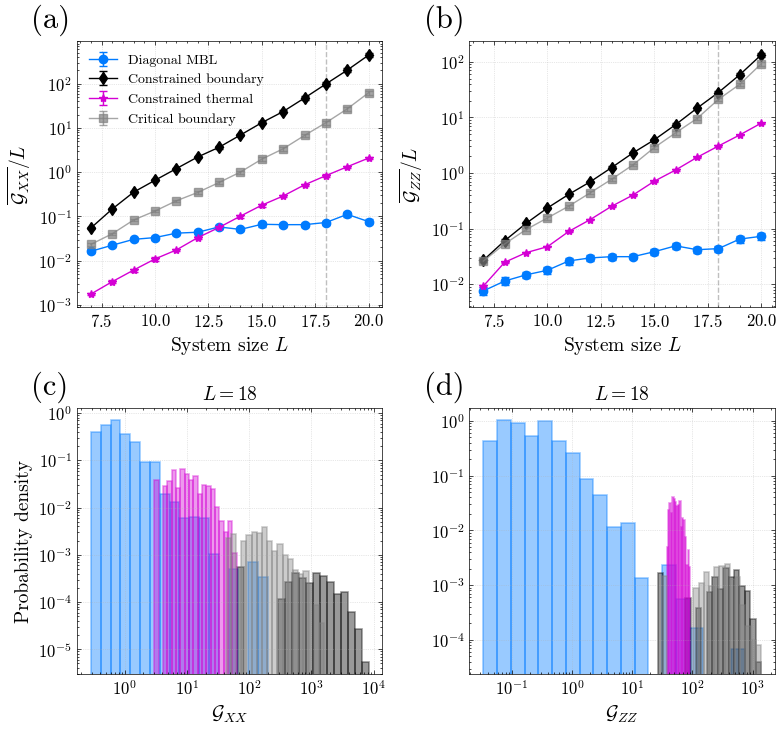}
    \caption{Scaling plot of the median HK metric components \textbf{(a,b)} for the PXP model and distribution of HK metric of the PXP Hamiltonian for $L=18$ with $100$ samples \textbf{(c,d)}. Note that the axes are on a logarithmic scale. We can identify that points in the constrained MBL phase (blue) experience a higher probability of outliers, whereas the others are more uniform (on the logarithmic scale).}
    \label{fig: pxp supp dist}
\end{figure}

In this subsection, we further analyze the phase diagram for other components (see Fig.~\ref{fig: ising supp}) and scaling of the Hilbert-Killing metric for different points in parameter space $(h_x/J, h_z/J)$. We see the results in Fig.~\ref{fig: ising more points}. We further provide the plot of the scaling behaviour of all components in Fig.~\ref{fig: ising supp scaling} for the Ising model. Furthermore, we show the distribution of a single phase space point in the PXP model in Fig.~\ref{fig: pxp supp dist}.
\subsection{Numerical values used}
In this section, we provide 3 tables that show the exact values used for the parameters in the scaling analyses in the main text and for Fig.~\ref{fig: ising more points} in the appendix. 
\begin{table}[htbp]
\centering
\begin{tabular}{c | c | c }
$h_x$ & $h_z$ & \texttt{matplotlib} Symbol \\ \hline 
0.01000000 & 1.11555556 & o  \\
0.80000000 & 0.00000000 & *  \\
0.23111111 & 0.45222222 & d  \\
2.00000000 & 1.11555556 & s  
\end{tabular}
\caption{Selected parameter values for $h_x$ and $h_z$ for Fig.~\ref{fig: ising}. Most values were selected from \texttt{numpy.linspace(0.01, 2, 10)}, which we used for the phase diagram sweep.}
\label{tab: ising params}
\end{table}
\begin{table}[htbp]
\centering
\begin{tabular}{c | c | c }
$h_x$ & $h_z$ & \texttt{matplotlib} Symbol \\ \hline 
1.33666667 & 1.33666667 & p  \\
0.45222222 & 0.45222222 & X  \\
0.67333333 & 0.45222222 & P  \\
0.01000000 & 0.01000000 & o  \\
0.01000000 & 1.11555556 & s  \\
1.33666667 & 0.67333333 & *  \\
2.00000000 & 1.11555556 & p  \\
1.11555556 & 0.23111111 & d  
\end{tabular}
\caption{Selected parameter values for $h_x$ and $h_z$ for Fig.~\ref{fig: ising more points}. Most values were selected from \texttt{numpy.linspace(0.01, 2, 10)}, which we used for the phase diagram sweep.}
\label{tab: ising more points parameters}
\end{table}
\begin{table}[htbp]
\centering
\begin{tabular}{c | c | c }
$W_z/W_x$ & $\lambda^0_x/W_x$ & \texttt{matplotlib} Symbol \\ \hline 
0.20 & 4.60 & o  \\
0.15 & 0.18 & d  \\
3.35 & 0.25 & *  \\
1.00 & 0.25 & s  
\end{tabular}
\caption{Selected parameter values for $W_z/W_x$ and $\lambda^0_x/W_x$ for Fig.~\ref{fig: pxp}.}
\label{tab: pxp params}
\end{table}

\subsection{Cutoff analysis}
\label{app: fid suz, adiabatic perturbation, linear response}
\begin{figure*}
    \centering
    \includegraphics[width=0.8\linewidth]{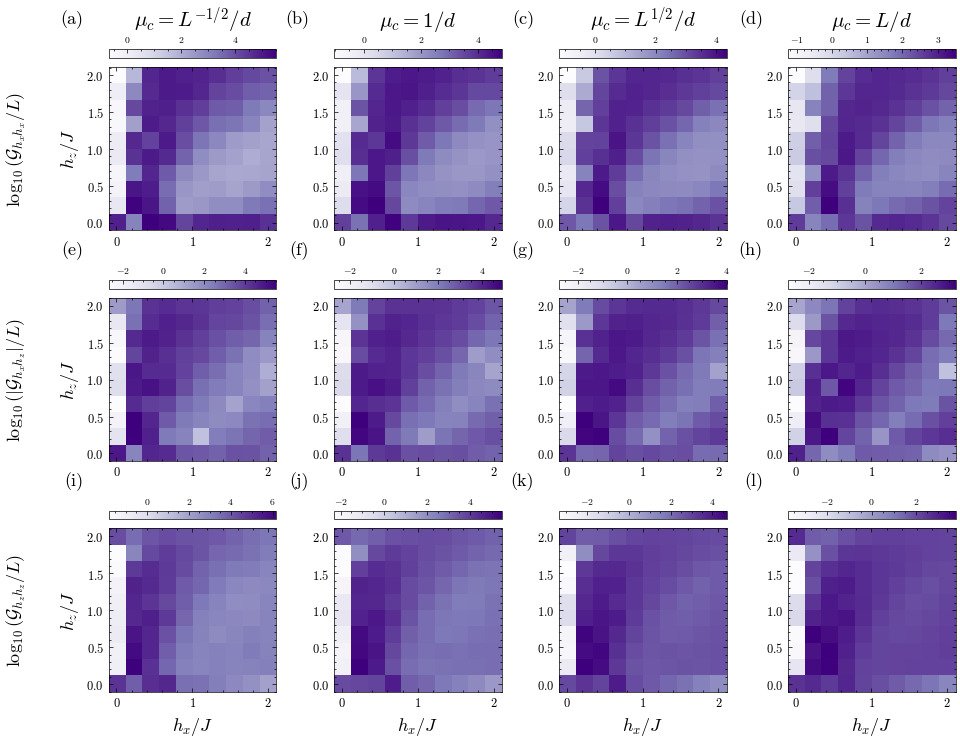}
    \caption{Cutoff analysis. We find that choosing a different cutoff $\mu_c=L^\alpha/d$, where $\alpha \in \{-1/2, 0, 1/2, 1\}$, does not alter the location of the boundary regime. The system size is $L=14$.}
    \label{fig: cutoff analysis}
\end{figure*}

The Hilbert-Killing metric is given by the eigenstate-averaged quantum metric tensor, which, in terms of the spectrum of the Hamiltonian for a given eigenstate $\ket{\psi_m}$, is given by
\begin{align}
    g_{\mu\nu}^{(m)}(\mu_c)=\Re \sum_{n\neq m}\frac{\mel{\psi_m}{\partial_\mu H}{\psi_n}\mel{\psi_n}{\partial_\nu H}{\psi_m}\omega_{nm}^2}{(\omega_{nm}^2+\mu_c^2)^2}.
\end{align}
where we define $\omega_{nm}=E_n-E_m$. Similar to~\cite{pandeyAdiabaticEigenstateDeformations2020, sharipovHilbertSpaceGeometry2026}, we introduce a cutoff $\mu_c=L/d$ to have numerical stability for near-degenerate eigenvalues. Our Hilbert-Killing metric is related to the standard quantum metric by an eigenstate averaging
\begin{align}
    \mathcal{G}_{\mu\nu}=\frac{1}{d}\sum_m g_{\mu\nu}^{(m)}(\mu_c),
\end{align}
In Fig.~\ref{fig: cutoff analysis} we analyze the results of the boundary regime as a function of different cutoffs $\mu_c=L^\alpha/d$, where $\alpha \in \{-1/2, 0, 1/2, 1\}$ and $\alpha=1$ is the one used in the main text. We find that the boundary regime is not altered as a function of the cutoff.
\begin{figure}
    \centering
    \includegraphics[width=\linewidth]{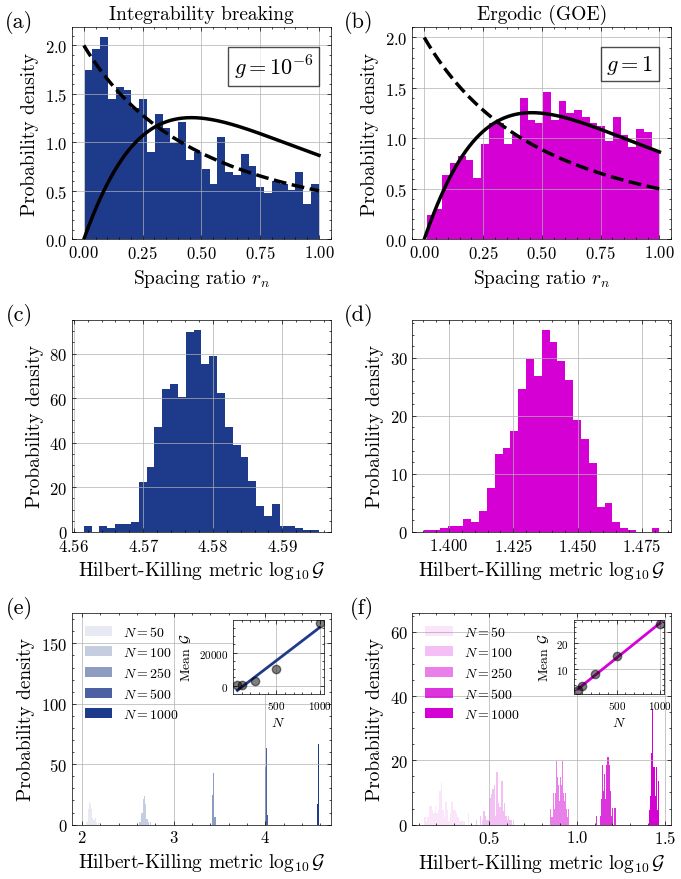}
    \caption{Scaling of HK metric with system size. Panels (a, b) show the distributions of the random matrix Hamiltonian for two values of the integrability-breaking parameter $g$ for a random Hamiltonian of size $N=1000$. The black lines correspond to the theoretical predictions for Poisson and GOE distributions~\cite{atasDistributionRatioConsecutive2013}. Panels (c, d) show the distributions of the Hilbert-Killing metric corresponding to the $N\times N$ random matrix. Note that the integrability-breaking one illustrates a larger response than the ergodic one. In (e, f) we inspect the system size scaling of the Hilbert-Killing metric. Both cases illustrate a scaling proportional to $N$, which would correspond to exponential scaling $N=2^L$ for a system of $L$ spins. We note, however, that the integrability-breaking case hosts an order-of-magnitude stronger response, as can be seen in the insets.}
    \label{fig: rmt vs N}
\end{figure}

\subsection{Random matrices and HK metric}
\label{app: spectral metric }

Random matrix models can serve as a proxy for studying the quantum phases of matter. Here, we aim to understand the interplay between random matrices and the HK metric. For that, we model a simple many-body Hamiltonian by
\begin{align}
    \hat{H}_\text{RMT}=\Lambda+g \hat{V}_\text{GOE},
\end{align}
where $\Lambda$ is a diagonal matrix that should capture Poissonian energy level statistics and $g$ captures the strength of the chaotic perturbation $\hat{V}_\text{GOE}$ from the Gaussian Orthogonal Ensemble (GOE). In practice, we sample the matrices $\Lambda, \hat{V}_\text{GOE}$ entries from uniform/normal distributions respectively~\footnote{Specifically, we sample from real matrices $A_{ij}\sim \mathcal{N}(0,1)$ and diagonalize $\Tilde{A}=(A+A^T)/2$ to get a orthogonal matrix.}, diagonalize the matrix for a fixed value of $g$, and normalize the resulting eigenvalues to lie in the interval $[-1,1]$ to avoid outliers from the normal distributions. This restricts the density of states to lie in the same interval. From the construction, we see that $g\to 0$ should tend toward Poissonian energy level statistics, and for finite $g$ it should mix and then be dominated by the GOE statistics. We illustrate the scaling of the HK metric as a function of the random matrix size $N$, for $g\neq 0$, in Figure~\ref{fig: rmt vs N}. We find that the mean value over the random samples increases, while the variance decreases. In addition, we find that the integrability-breaking case shows a stronger response overall.

\newpage

\bibliography{references}

\end{document}